\documentstyle[array,multicol,prb,aps,psfig]{revtex}
\draft
\tighten
\begin{document}

\title{Quasiballistic correction to the density of states in
three-dimensional metal}
\author{A.~ A.~ Koulakov}
\address{Theoretical Physics Institute, University of Minnesota,
Minneapolis, Minnesota 55455}

\date{\today}

\maketitle
\begin{abstract}
We study the exchange correction to the density of states in the
three-dimensional metal near the Fermi energy. 
In the ballistic limit, when the distance to the Fermi level
exceeds the inverse transport relaxation time $1/\tau$, we find the
correction linear in the distance from the Fermi level. By a large
parameter $\epsilon_{\rm F} \tau$ this ballistic correction exceeds the 
diffusive correction obtained earlier.
\end{abstract}
\begin{multicols}{2}

The zero-bias tunneling anomaly in disordered metals has been
studied extensively both experimentally\cite{Experiment,Boris} and 
theoretically\cite{AA85,Rudin,Boris}. The explanation of this phenomenon has
been given on the basis of the interaction induced correction
to the single-particle density of states (DOS). 
For the three-dimensional case the leading order exchange correction is given by:
\begin{equation}
\frac{\delta \nu_{\rm diff}}{\nu_0} = A \frac{\lambda}
{\left(\epsilon_{\rm F}\tau\right)^2} \sqrt{|\varepsilon| \tau}.
\label{diff_corr}
\end{equation}
Here $\nu_0=m^2v_{\rm F}/\pi^2$ is DOS on the Fermi level in the
non-interacting metal ($m$ and $v_{\rm F}$ being the mass of electron and the Fermi velocity correspondingly, $\hbar=1$), 
$A=3^{3/2}/8\sqrt{2}\approx 0.459$ is the numerical constant, 
and $\lambda$ is the unitless interaction strength. In the realistic cases $\lambda \approx 1$. 
The subscript attached to the correction to 
DOS $\delta\nu_{\rm diff}$ emphasizes that the above result is valid
in the diffusive limit, i. e. when the distance to the Fermi level $|\varepsilon|$ 
is much smaller than the inverse transport scattering time $1/\tau$. The
maximum value that the correction can reach can therefore be estimated as
$\delta\nu / \nu \lesssim (\epsilon_{\rm F} \tau )^{-2} \ll 1$.


In this work we evaluate the exchange correction to 
DOS in three-dimensional metal with short-range impurities. 
The use of the concrete form of disorder allows us to 
go beyond the universal diffusive regime.
Our result in the ballistic regime ($|\varepsilon| \gg 1/\tau$) is:
\begin{equation}
\frac{\delta \nu_{\rm ball}}{\nu_0} = B \frac{\lambda}
{\left(\epsilon_{\rm F}\tau\right)^2} |\varepsilon| \tau.
\label{ball_corr}
\end{equation}
Here $B = \pi/16 \approx 0.196$ is the numerical constant, 
$\lambda = \tilde{V}(\omega=0, \bbox{q}=0)\nu_0$, where $\tilde{V}(\omega, \bbox{q})$ 
is the Fourier transform of screened electron-electron interaction potential. This result
is valid up until energies of the order of Fermi energy. 
The maximum value reached by this correction can be estimated as 
$\delta\nu / \nu \lesssim (\epsilon_{\rm F} \tau )^{-1}$. 
Thus it exceeds the diffusive correction by a large parameter $\epsilon_{\rm F} \tau$.
We conclude therefore that the ballistic correction produces a larger suppression
of DOS on the Fermi level than the diffusive correction in case of short range
impurities.
 
The singular diffusive correction to the tunneling DOS is observed in many experiments.\cite{AA85,Boris} 
However at larger energies it crosses over into
less singular correction, behaving approximately as the absolute value
of distance to the Fermi level. This behavior is consistent with 
our prediction (\ref{ball_corr}) for the ballistic regime. 
The details of the crossover between (\ref{diff_corr}) and (\ref{ball_corr}) are given below.

To derive Eq.~(\ref{ball_corr}) we follow the guidelines of Refs.~\onlinecite{AA85} and \onlinecite{Rudin}.
The correction to one particle DOS is related to the exchange correction to the retarded Green function
\begin{equation}
\delta \nu(\varepsilon) = -\frac{2}{\pi} \int \frac{d\bbox{p}}{(2\pi)^3} {\rm Im} \delta G^R(\varepsilon, \bbox{p})
\end{equation}
The latter is calculated in the first order of the perturbation theory in the electron-electron interaction
$\tilde{V}(\omega, \bbox{q})$ 
\begin{equation}
\begin{array}{l}
{\displaystyle \delta G^R(\varepsilon, \bbox{p}) = i \left[ G^R(\varepsilon, \bbox{p})\right]^2 \int \frac{d\bbox{q}}{(2\pi)^3} 
\frac{d\omega}{2\pi} } \\ \\
{\displaystyle \times  \left[ \Gamma(\omega, \bbox{q})^2 - 1\right]
G^A(\varepsilon-\omega, \bbox{p-q})\tilde{V}(\omega, \bbox{q}).}
\end{array} 
\label{def_corr}
\end{equation}
Here $G^R$ and $G^A$ are the retarded and advanced Green functions respectively 
$G^{R, A}(\bbox{p}, \varepsilon) = 1/\left(\varepsilon - \bbox{p}^2/2m \pm i/2\tau \right)$. 
We subtract unity from the square of the vertex function  
in the presence of impurities $\Gamma(\omega, \bbox{q})$ to exclude ``bare'' interaction 
correction existing with no impurities. The vertex function has to be calculated 
using our model of impurity potential
\begin{equation}
u(\bbox{r}) = \sum_i u_0 \delta (\bbox{r} - \bbox{r}_i),
\end{equation} 
where the locations of impurities $\bbox{r}_i$ are scattered randomly with average density $n_i$.
In the ladder approximation the vertex is then given by
\begin{equation}                         
\begin{array}{l}
{\displaystyle \Gamma(\omega, \bbox{q}) = \theta \left[\varepsilon(\varepsilon-\omega)\right]
+ \frac{\theta (\varepsilon)\theta (\omega-\varepsilon) }{ 1- \zeta(\omega, \bbox{q})}} \\ \\
{\displaystyle +
\frac{\theta (-\varepsilon)\theta (\varepsilon-\omega)}{ 1- \zeta^*(\omega, \bbox{q})},}
\end{array}
\end{equation}
where
\begin{equation}                         
\begin{array}{l}
{\displaystyle \zeta(\omega, \bbox{q}) = n_i |u_0|^2 \int \frac{d\bbox{p}}{(2\pi)^3} G^R(\bbox{p}+\bbox{q}, \varepsilon) 
G^A(\bbox{p}, \varepsilon-\omega)} \\ \\
{\displaystyle \ \ \ = \frac {i}{2qv_{\rm F}\tau} \ln \frac{\omega+qv_{\rm F} +i/\tau}{\omega - qv_{\rm F} +i/\tau} },
\end{array}
\end{equation}                                                                                                     
and $1/\tau = \pi\nu_0 n_i|u_0|^2$. The latter expression is valid for any $\omega,\ \varepsilon,\ qv_{\rm F} \ll \epsilon_{\rm F}$.
In the diffusive case ($\omega,\ \varepsilon,\ qv_{\rm F} \ll 1/\tau$) the last expression takes the form
$\zeta = 1+i\omega\tau - (qv_{\rm F}\tau)^2/3$ and the vertex part has the expected diffusive pole.
 
The screened electron-electron interaction $\tilde{V}(\omega, \bbox{q}) = 4\pi e^2/\left[q^2+4\pi e^2\Pi(\omega, \bbox{q})\right]$,
where the polarization operator derived in the random phase approximation
\begin{equation}
\Pi(\omega, \bbox{q}) = \nu_0\left[ 1 + \alpha i\omega \tau \zeta / (1-\zeta)\right].
\end{equation}
Here parameter $\alpha$ is equal to 1 or 0 depending on whether the retardation of the interaction
is to be taken into account or not. This parameter is introduced for comparison to the earlier results.\cite{AA85}

After some transformations the expression for the correction to DOS is obtained from Eq.~(\ref{def_corr})
\begin{equation}
\frac{\delta \nu}{\nu_0} = \frac {\lambda}{(\epsilon_{\rm F}\tau)^2} f\left(|\varepsilon|\tau\right), 
\end{equation} 
where the interaction constant $\lambda=1$ and
\begin{equation}
\begin{array}{l}
{\displaystyle f( \gamma ) = \frac{1}{8\pi} \int_0^\gamma d\gamma' {\rm Im} \int_0^{\infty}  \frac{x^2 dx}{x^2 - (\gamma'+i)^2}} \\ \\
{\displaystyle \times   \frac {2\zeta- \zeta^2} {(1-\zeta)\left[ 1-\zeta(1-\alpha i\gamma') \right]} ,} \\ \\ 
{\displaystyle \zeta = \frac {i}{2x} \ln \frac{\gamma'+x+i}{\gamma'-x+i}}.
\end{array}   
\label{answer_int}
\end{equation}    
The integrals in (\ref{answer_int}) cannot be evaluated in terms of elementary functions. Nevertheless the
asymptotic expressions for diffusive ($\gamma\ll 1$) and ballistic ($\gamma \gg 1$) cases are easily obtained:
\begin{equation}            
f(\gamma) = \left\{
\begin{array}{ll}
{A\sqrt{\gamma}, }&{\gamma \ll 1} \\ 
{B\gamma, }&{\gamma \gg 1.}
\end{array}        
\right.
\end{equation}                                                                                                
They lead to Eqs.~(\ref{diff_corr}) and (\ref{ball_corr}). The constant $A$ in this formula differs
for the cases of instantaneous ($\alpha=0$) and retarded ($\alpha=1$) interactions. For the former case
we obtain $A=3^{3/2}/16\sqrt{2}$, for the latter $A=3^{3/2}/8\sqrt{2}$. This agrees with the previous results.\cite{AA85} 
The ballistic constant $B=\pi/16$ is the same for both instantaneous and retarded cases.

\begin{figure}
\centerline{
\psfig{file=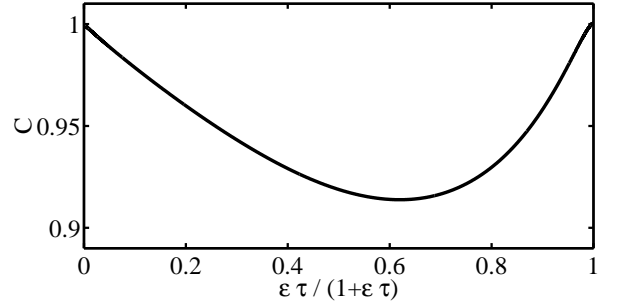,width=3.0in,bbllx=34pt,bblly=220pt,bburx=538pt,bbury=484pt}
}
\setlength{\columnwidth}{3.2in}
\centerline{\caption{The crossover function.
\label{fig10}
}}
\vspace{-0.1in}
\end{figure}

The crossover between diffusive and ballistic regimes can be described numerically. To this end we
represent the correction to DOS in the intermediate region as follows
\begin{equation}
\delta \nu = C \sqrt{\delta \nu_{\rm diff}^2+\delta \nu_{\rm ball}^2},
\end{equation}
where the crossover function $C\approx 1$ is shown in Fig.~\ref{fig10}.
%
%
The crossover function can be effectively fitted with polynomials
\begin{equation}
C(\xi) = \sum_{n} C_n \xi^n,\ \xi \equiv \varepsilon \tau / \left(1+\varepsilon \tau\right).
\end{equation}
The fit with a $7$th degree polynomial with coefficients 
$C_0=0.9999$,
$C_1=-0.2610$,
$C_2=0.9114$,
$C_3=-5.6062$,
$C_4=17.9270$,
$C_5=-28.7905$,
$C_6=22.8587$, and
$C_7=-7.0387$ guarantees an error not exceeding $0.14\%$.

In conclusion we evaluated the exchange correction to the tunneling density of states in three-dimensional metal
in the ballistic regime ($1/\tau \ll \varepsilon \ll \epsilon_{\rm F}$). The obtained correction
is proportional to the distance to the Fermi level $\varepsilon$ and exceeds the diffusive correction
by the large parameter $\epsilon_{\rm F}\tau$. The crossover between diffusive and ballistic limits is
also studied.

The author is grateful to Boris Shklovskii and Alexander Rudin for valuable discussions.
This work is supported by NSF grant DMR-9616880.


\end{multicols}
\end{document}